\documentclass[fleqn,twoside]{article}
\usepackage{amssymb}
\usepackage{graphicx}
\usepackage{espcrc2}

\def\mbf#1{\hbox{\boldmath$#1$}}
\def\embf#1{\hbox{\scriptsize\boldmath$#1$}}
\newcommand{\bDelta}{\mbf\Delta}
\newcommand{\bdelta}{\mbf\delta}

\def\pom{\hbox{$\mathcal P$}}
\def\oddindex{{\hbox{\scriptsize$\mathcal O$}}}

\def\eq{{}={}}

\newif\ifhepph
\hepphtrue

\begin{document}

\title{
\ifhepph\vbox to 0pt{{\vss\flushright\normalsize\rm
HD-THEP-02-27\\
hep-ph/0208175\\
\endflushright}}\fi
The perturbative odderon in elastic $pp$ and $p\bar p$ scattering}

\author{V.~Schatz
\address{Institut f\"ur Theoretische Physik, Universit\"at Heidelberg, \\
Philosophenweg 16, 69120 Heidelberg, Germany}}

\begin{abstract}
Different models for the odderon-proton coupling are considered and their
effects on the differential cross section in the dip region in elastic $pp$ and
$p\bar p$ scattering are investigated.  An allowed range for the size of a
possible diquark cluster in the proton can be obtained from a geometrical
model.
\end{abstract}

\maketitle

\section{Introduction}

The evidence for the existance of an odderon~\cite{luk} remains scarce.  To
date, the best evidence comes from the differential cross section in elastic
$pp$ and $p\bar p$ scattering.  The $pp$ data show a dip structure at
$-t\approx1.3\;$GeV$^2$ while the $p\bar p$ data only flatten off at that
point.  The odderon, being odd under charge conjugation, couples with different
signs to protons and antiprotons and can hence account for that difference.

We describe the odderon in leading order perturbation theory, as a perturbative
three-gluon exchange in a $C=-1$ state.  We assume that the scale
$1.3\;$GeV$^2$ is large enough for perturbation theory to be valid; and we do
no $\log s$ resummation.  We consider three different models for the
odderon-proton coupling: one proposed by Fukugita and Kwieci\'nski, one
proposed by Levin and Ryskin and a geometrical model which allows an estimation
of the size of a possible diquark cluster in the proton.

\section{Method of calculations}

\subsection{The framework}

The data for the differential cross section in $pp$ and $p\bar p$ elastic
scattering are well described by the Donnachie-Landshoff (DL) fit~\cite{dlfit}.
The authors use a number of contributions: Pomeron (\pom), Reggeon, triple
gluon (odderon), \pom\pom, \pom\pom\pom, \pom+Reggeon, and \pom+double gluon
exchange.  Their perturbative triple-gluon exchange contribution is charge
conjugation odd due to the colour structure of the single Feynman graph taken
into account, hence an odderon.  

We use the DL fit as a framework for comparing different odderon contributions
to experimental data.  To that end we replace their triple-gluon exchange
amplitude by one of the model odderon contributions.  We retain the original
parameter values of the fit and make no attempt to improve it.

\subsection{A position-space model}

Our prime interest was to investigate the influence of the proton structure on
the odderon-exchange amplitude and hence on the differential cross section.
The scattering amplitude in position space is given by
\begin{eqnarray}
T_\oddindex(s,t) &\eq& 2 \, s \int d^2 \mbf b \; e^{-i\embf q\embf b}
                     \int d^2 \mbf R_1 \int d^2 \mbf R_2 \\
&&                {}\times|\psi(\mbf R_1)|^2 \; |\psi(\mbf R_2)|^2 \; 
                     J(\mbf b, \mbf R_1, \mbf R_2)\,.\nonumber
\end{eqnarray}
The two latter integrations are over the size and orientation of the protons in
transverse position space.  $\psi$ is the proton wave funtion.  $J=S-1$ is the
reduced scattering amplitude or $T$-Matrix element.  It is computed with a
method developed by Nachtmann~\cite{nacht} based on the functional
representation of scattering matrix elements and the WKB approximation.  For a
complete presentation of this method, please refer to~\cite{nacht}.

In our case, it leads to a correlator of six integrals over gluon fields along
the paths of the quarks.  We then project out the $C=-1$ part to obtain the
odderon.  See~\cite{des} for more details.

We use a Gaussian wave function for the proton,
\begin{equation}
\psi(\mbf R)=\sqrt\frac2\pi\,\frac1S\,
                 \exp\!\left(-\frac{|\mbf R|^2}{S^2}\right)\,.
\end{equation}
The parameter $S$ is the proton size, which we set to 0.8$\;$fm.

\begin{figure}
\center\includegraphics[height=2cm]{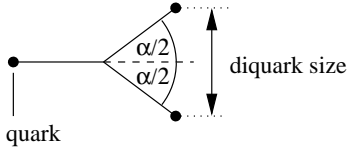}\endcenter
\label{fig:starmodel}
\caption{The geometric model for the proton}
\end{figure}

In addition to the wave function we make a model for the position of the quarks
inside the proton.  Two of the quarks form a diquark cluster as shown in
Fig.~\ref{fig:starmodel}.  The diquark size is a free parameter.  By comparing
with experiment, we will be able to place a bound on it.

\subsection{Momentum space impact factors}

Two impact factors we took from the literature are defined in momentum space.
In momentum space, the odderon exchange amplitude is computed by folding two
odderon-proton impact factors with the propagators of the three gluons which
make up the odderon:
\begin{eqnarray}
T_\oddindex(s,t) &\eq& \frac s{32} \frac56 \frac1{3!} \int
        \frac{d^2\bdelta_{1t}}{(2\pi)^2} \frac{d^2\bdelta_{2t}}{(2\pi)^2} \\
&& \hskip5mm{}\times\frac{\Phi_p^2(\bdelta_{1t},\bdelta_{2t},\bDelta_t)}
        {\bdelta_{1t}^2\, \bdelta_{2t}^2\, 
                (\bDelta_t-\bdelta_{1t}-\bdelta_{2t})^2} \,.\nonumber
\end{eqnarray}
The $\bdelta_{it}$ are the transverse gluon momenta;
$\bDelta_t=\bdelta_{1t}+\bdelta_{2t}+\bdelta_{3t}$ is the transverse momentum
of the odderon.

For reasons of gauge invariance the impact factor has to be of the form
\begin{eqnarray}
&&\Phi_p(\bdelta_{1t}, \bdelta_{2t}, \bDelta_t) =
        8\,(2\pi)^2\, g^3\, \Bigg[ F(\bDelta_t,0,0) \\
&& \kern 5mm {} - \sum_{i=1}^3 F(\bdelta_{it},\bDelta_t-\bdelta_{it},0) 
     + 2 F(\bdelta_{1t}, \bdelta_{2t}, \bdelta_{3t}) \Bigg]\,, \nonumber
\end{eqnarray}
where $F$ is a form factor.

One of the form factors we used was proposed by Fukugita and
Kwieci\'nski~\cite{fk}:
\begin{equation}
F_{FK}(\bdelta_{1t}, \bdelta_{2t}, \bdelta_{3t}) = \frac{A^2}{A^2 +
           \frac12 \sum\limits_{i\ne k}(\bdelta_{it}-\bdelta_{kt})^2 }\,.
\end{equation}
The constant $A$ determines the width of the form factor and equals half the
rho mass.  The other form factor was published by Levin and Ryskin~\cite{lr}:
\begin{equation}
F_{LR}(\bdelta_{1t}, \bdelta_{2t}, \bdelta_{3t}) = \exp\left(
                -R^2\, \sum_{i=1}^3 \bdelta_{it}^2 \right)\,.
\end{equation}
$R=0.33\;$fm is the proton radius used by the authors. We did not attempt to
fit the parameters of either form factor but kept the values supplied by the
authors.

\section{Results}

\begin{figure*}[t]
\moveright 22mm\vbox{\input{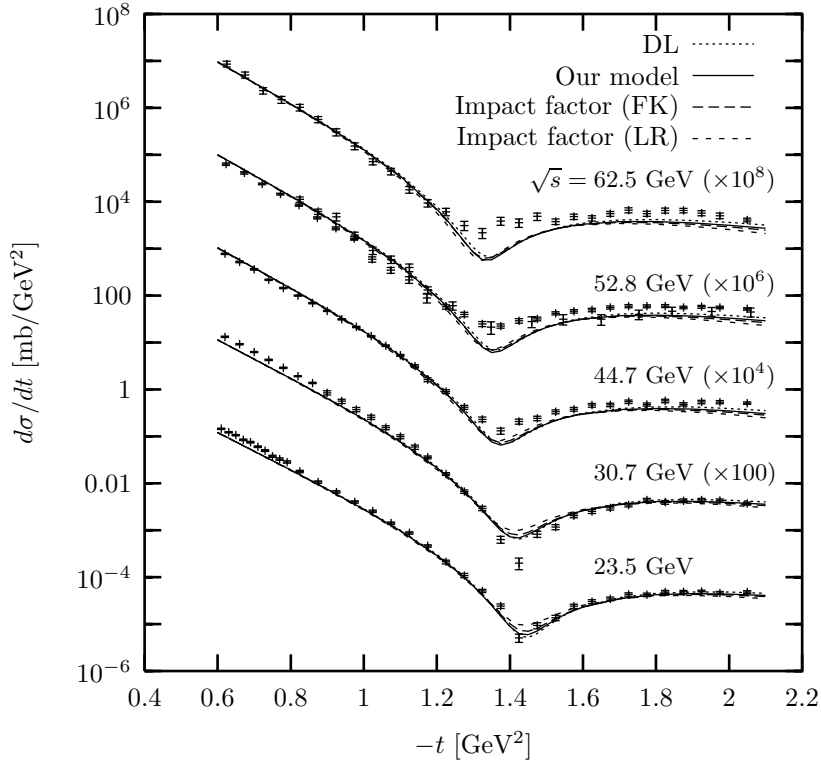}}
\caption{Differential elastic cross section for $pp$ scattering. All three
models for the odderon-proton coupling and the original Donnachie-Landshoff fit
are compared with experimental data~\cite{ama,break}.  All five ISR energies
are displayed, with successive energies shifted upwards by a factor of 100.
The data do not favour one model over the others.}
\label{fig:pp}
\end{figure*}

\begin{figure*}
\moveright 35mm\vbox{\input{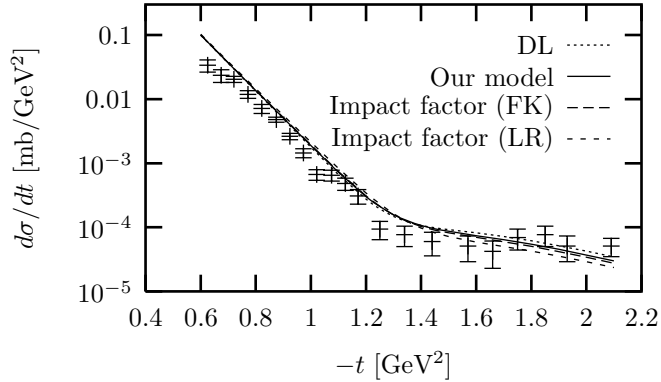}}
\caption{Differential elastic cross section for $p\bar p$ scattering for all
models for the odderon-proton coupling and the original Donnachie-Landshoff fit
compared with experimental data~\cite{break}.  The centre-of-mass energy is
$\sqrt s=53\;$GeV.  Again, the data do not favour one model over the others.}
\label{fig:ppbar}
\end{figure*}

As can be seen from Figs.\ \ref{fig:pp} and~\ref{fig:ppbar}, all models provide
a satisfactory description of the data provided their parameters are adjusted
correctly.  The data do not favour one model over the others.

In the case of the geometrical model, there are two parameters, the diquark
size and the coupling constant.  Since the precise value of the coupling
constant is unknown in a leading-order calculation, we fitted the diquark size
for several values of $\alpha_s$ which are common in the literature at the
scale given by $-t\approx1.3\;$GeV$^2$ in the dip region.  

Table~\ref{tab:rdiq} shows the results.  For $\alpha_s\geq0.3$ the diquark size
is $\lesssim0.35\;$fm.  This result is of great importance for nonperturbative
calculations where the proton is often described as a colour dipole.  For such
a small diquark size this is legitimate since soft gluons cannot resolve the
diquark.

\begin{table}
\vbox{
\offinterlineskip
\tabskip=0pt
\halign{ 
\vrule height2.5ex depth1.0ex
#\tabskip=0.5em & \hfil #\hfil &\vrule # & \ $#\,\pi$\hfil &\vrule # &
                 \hfil # \hfil &#\vrule\tabskip=0pt \cr\noalign{\hrule}
& $\alpha_s$ &&\omit angle $\alpha$ && mean diquark size [fm] & \cr\noalign{\hrule}
& 0.3 && 0.22 && 0.34 &\cr
& 0.4 && 0.14 && 0.22 &\cr
& 0.5 && 0.095 && 0.15 &\cr\noalign{\hrule}
}}
\caption{Best fit values for the diquark size}
\label{tab:rdiq}
\end{table}

In the calculations with momentum space impact factors, the only free parameter
is the coupling constant.  The best fit was $\alpha_s=0.3$ for the
Fukugita-Kwieci\'nski (FK) form factor and $\alpha_s=0.5$ for the Levin-Ryskin
form factor.  With a larger value for the coupling constant, eg $\alpha_s=1$,
the curves would overshoot the data by more than an order of magnitude over
nearly the whole $t$ range.

This casts some doubt on predictions of diffrative $\eta_c$ production.  Three
groups \cite{kwie_eta,eng_eta,bart_eta} have used the FK impact factor with a
value of $\alpha_s=1$ in calculations of the diffractive $\eta_c$ production
amplitude.  In view of our results, this looks like a significant
overestimation.

\section*{Acknowledgements}

I would like to thank my collaborators H.~G.~Dosch and C.~Ewerz for their
contributions to the results presented here.

\section*{Discussion}

{
\parindent=0pt

{\bf M.~Boutemeur} (Munich){\bf:} {\it Your model does not describe the $-t$
distributions at higher energies.  Do you have in mind other ingredients to
your model to make it fit the data better?}

{\bf V.~Schatz:} The DL fit is less good at higher energies, but the odderon
alone cannot remedy that.  Due to the uncertainties in our calculations---the
unkown exact value of the coupling, the possibility of non-perturbative effects
playing a role---we were not interested in a precision fit. There is a better
fit due to Gauron, Leader and Nicolescu, but it is incompatible with the
odderon contributions we investigated.

{\bf L.~Le\'sniak} (Krakow){\bf:} {\it The momentum transfer distribution
$d\sigma/dt$ in $pp$ or $p\bar p$ scattering can depend on five different spin
amplitudes. Is the spin dependence of the $pp$ or $p\bar p$ amplitudes included
in your model or do you use only one spin independent amplitude?}

{\bf V.~Schatz:} The different spin amplitudes are contained in our amplitudes
for the various contributions. However, we did not try to extract or
investigate the interplay of different spin amplitudes.
}

\end{document}